\documentclass{aa}

\usepackage{graphicx}
\usepackage{amsmath,epsfig,rotating,amssymb}

\begin{document}

\def \rot{{\rm {\bf rot} }}
\def \grad{{\rm {\bf grad} }}
\def \div{{\rm div}}
\def \cha{\widehat}
\def \pr{{\it permanent}  regime }


\author{ Hennebelle P. \inst{1} and Ciardi A. \inst{1} \inst{2}}

\institute{ Laboratoire de radioastronomie, UMR 8112 du CNRS, 
 {\'E}cole normale sup{\'e}rieure et Observatoire de Paris, 24 rue Lhomond,
 75231 Paris cedex 05,
France \and
 Laboratoire de Physique des Plasmas, UMR 7648 du CNRS, 10-12 Avenue de l'Europe, Velizy, 78140, France }

\offprints{ P. Hennebelle  \\
{\it e-mail:} patrick.hennebelle@ens.fr}   

\title{Disk formation during collapse of magnetized protostellar cores}

\titlerunning{Disk formation during collapse}

\abstract{In the context of star and planet formation, understanding the formation of disks is of fundamental importance.}
{Previous studies found that the magnetic field has a very strong impact on the collapse of 
a prestellar cloud, by possibly suppressing the formation of a disk even for relatively
modest values of the magnetic intensity. Since observations infer  that cores have a substantial 
level of magnetization, this raises the question of how disks form. However, most  studies
have been restricted to the case in which the initial angle, $\alpha$, between the magnetic field 
and the rotation axis  equals 0$^\circ$. Here we explore and analyse the influence of 
non aligned configurations on  disk formation.} 
{We perform 3D ideal MHD, AMR numerical simulations for various values of $\mu$, the ratio of the 
mass-to-flux to the  critical mass-to-flux, and various values of $\alpha$.} 
{We find that disks form more easily as $\alpha$  increases from 0 to 90$^\circ$. 
We propose that as the magnetized pseudo-disks become thicker with increasing $\alpha$, the magnetic 
braking efficiency is lowered. We also find that even small values of $\alpha$ ($\simeq$ 10-20$^\circ$) 
show significant differences with the aligned case.}  
{Within the framework of ideal MHD, and for our choice of initial conditions, centrifugally supported disks 
cannot form for values of $\mu$ smaller than $\simeq$3 when the magnetic field and the rotation axis 
are perpendicular, and smaller than about $\simeq$5-10  when they are perfectly aligned.  }

\maketitle

\section{Introduction}
Understanding the formation of protostellar disks is an important issue in the 
context of star and planet formation. The presence of circumstellar 
disks is well established around T-Tauri stars (Watson et al. 2007), as well as around younger class I and II protostars. 
However, because of difficulties in disentangling the emission of the disk and envelope,
the existence of disks around class 0 objects 
remains a matter of debate (Mundy et al. 2000, Belloche et al. 2002, Jorgensen et al. 2007).

Starting with a level of rotation consistent with observations, hydrodynamical numerical simulations 
discovered that  centrifugally supported disks form with characteristic diameters 
of the order of a few hundred AU 
(Matsumoto \& Hanawa 2003, Hennebelle et al. 2004, Goodwin et al. 2007, Commer{\c c}on et al. 2008). These structures 
are the unavoidable consequence of angular momentum conservation during the collapse phase. 
In addition, since these disks are massive 
and strongly self-gravitating, they  are prone to gravitational instability and have often been 
found to fragment, leading to the formation of a small cluster of stars.

However,  magnetic fields appear to have an important impact. Magnetohydrodynamic (MHD) simulations 
(e.g.; Machida et al. 2005, Banerjee \& Pudritz 2006) 
find that even for modest values of the 
magnetic intensity, disk formation can be suppressed by magnetic braking 
 (Shu et al. 1987, Mouschovias 1991, Basu \& Mouschovias 1995, Galli et al. 2006)
 which transports angular momentum from the inner parts of the cloud 
towards its outer regions (Allen et al. 2003, Fromang et al. 2006, 
Price \& Bate 2007, Hennebelle \& Fromang 2008 (hereafter HF08), Mellon \& Li 2008, 2009).

The typical values of $\mu$, the ratio of the mass-to-flux to the critical 
mass-to-flux ratio, at which the  formation of a rotationally supported disk is suppressed, 
vary from one study to another; probably because of different choices of 
initial conditions. For example, while Mellon \& Li (2008) estimate that the value of $\mu$, below which disks do
not form,  is larger than 10, Price \& Bate (2007) and HF08 find that this 
occurs when $\mu <  5-10$.
A  broad distribution 
of $\mu$ has been inferred from observations, but most cores have magnetic fields  corresponding to values
 of $\mu$ typically smaller than 5 (Crutcher 1999). 
This  raises  a fundamental question about how disks can form in spite of the relatively
high degree of magnetization.

Most theoretical studies 
have been  performed by assuming a simple initial configuration
in which the magnetic field and the rotation axis are parallel. 
Notable exceptions are the works of
Machida et al. (2006), who explored the influence of a non zero angle between the magnetic field axis 
and the rotation axis, and Price \& Bate (2007) who considered the case where the 
two axes are perpendicular. 
However, none of these works  focused on the problem of disk formation and magnetic 
braking in great details, which are the topics of the present work.

The paper is organised as follows: in sect.~2, we  analyse the magnetic braking  emphasizing the 
importance of the initial angle, $\alpha$; in sect.~3 numerical simulations
of  collapsing cores for various values of $\alpha$ and $\mu$ are presented; sect.~4 concludes the paper.

\setlength{\unitlength}{1cm}
\begin{figure*}[t]
\begin{picture}(0,10.5)
\put(0,5.5){\includegraphics[width=5.5cm]{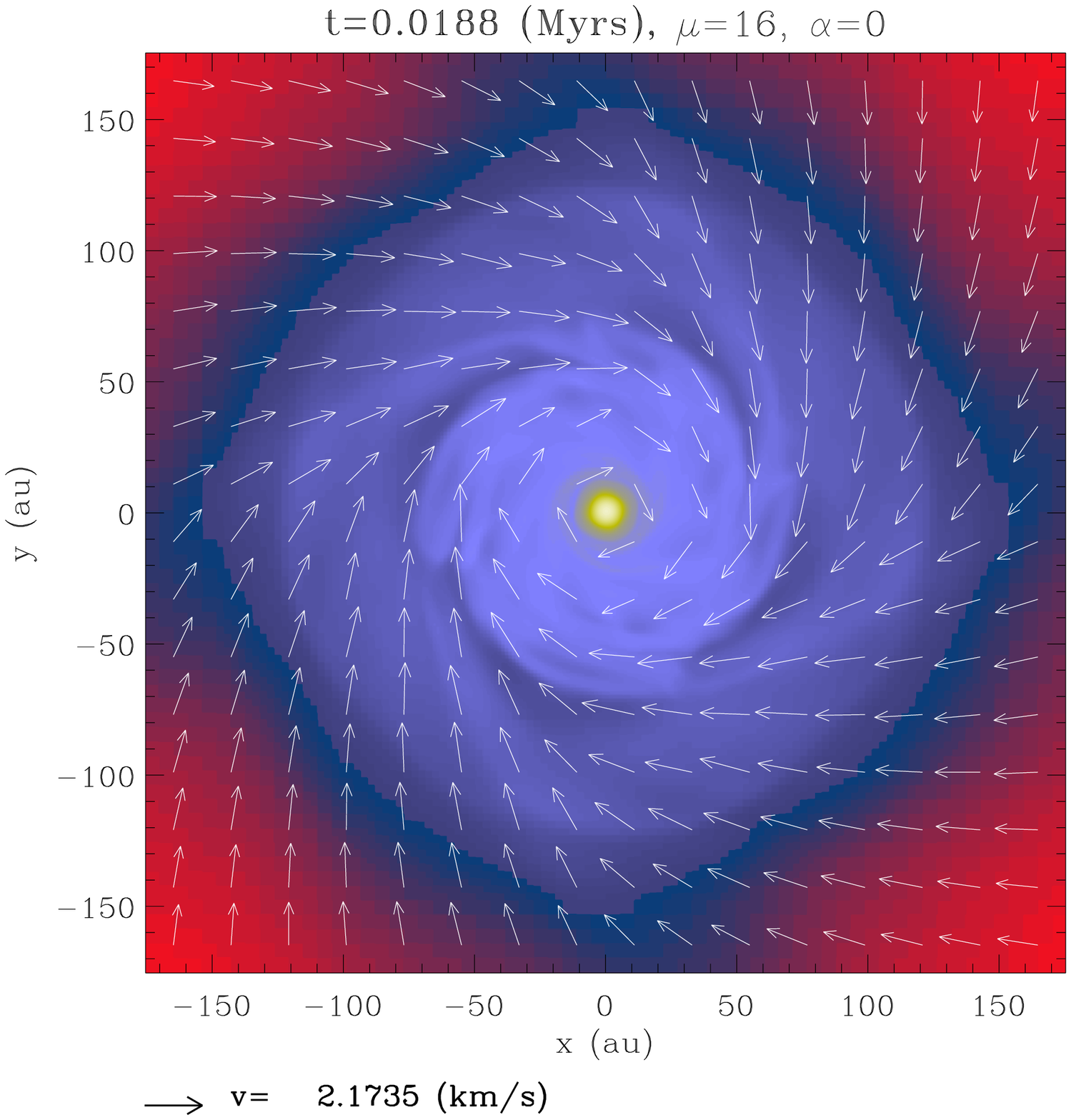}}
\put(0,0){\includegraphics[width=5.5cm]{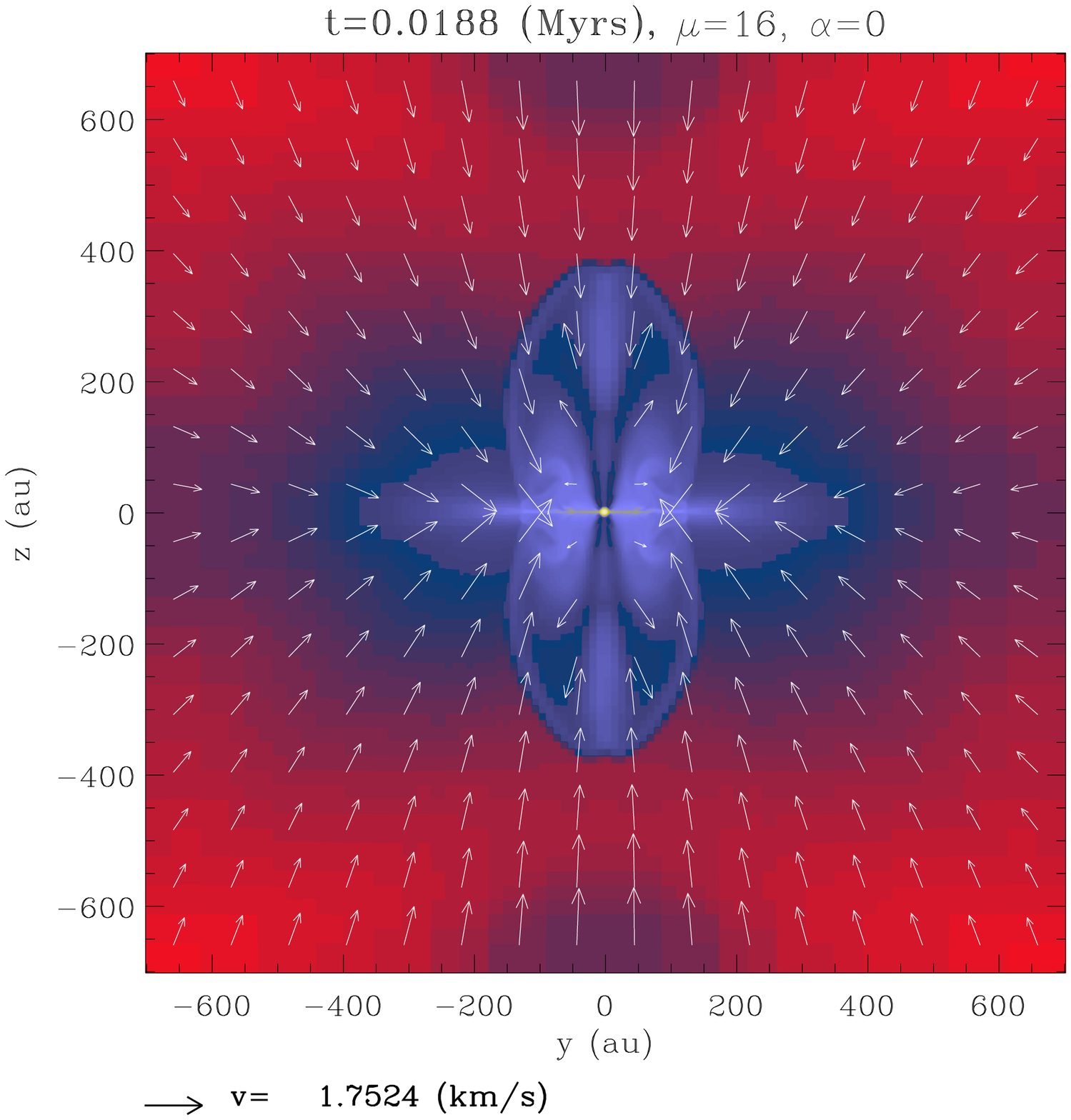}}
\put(6,5.5){\includegraphics[width=5.5cm]{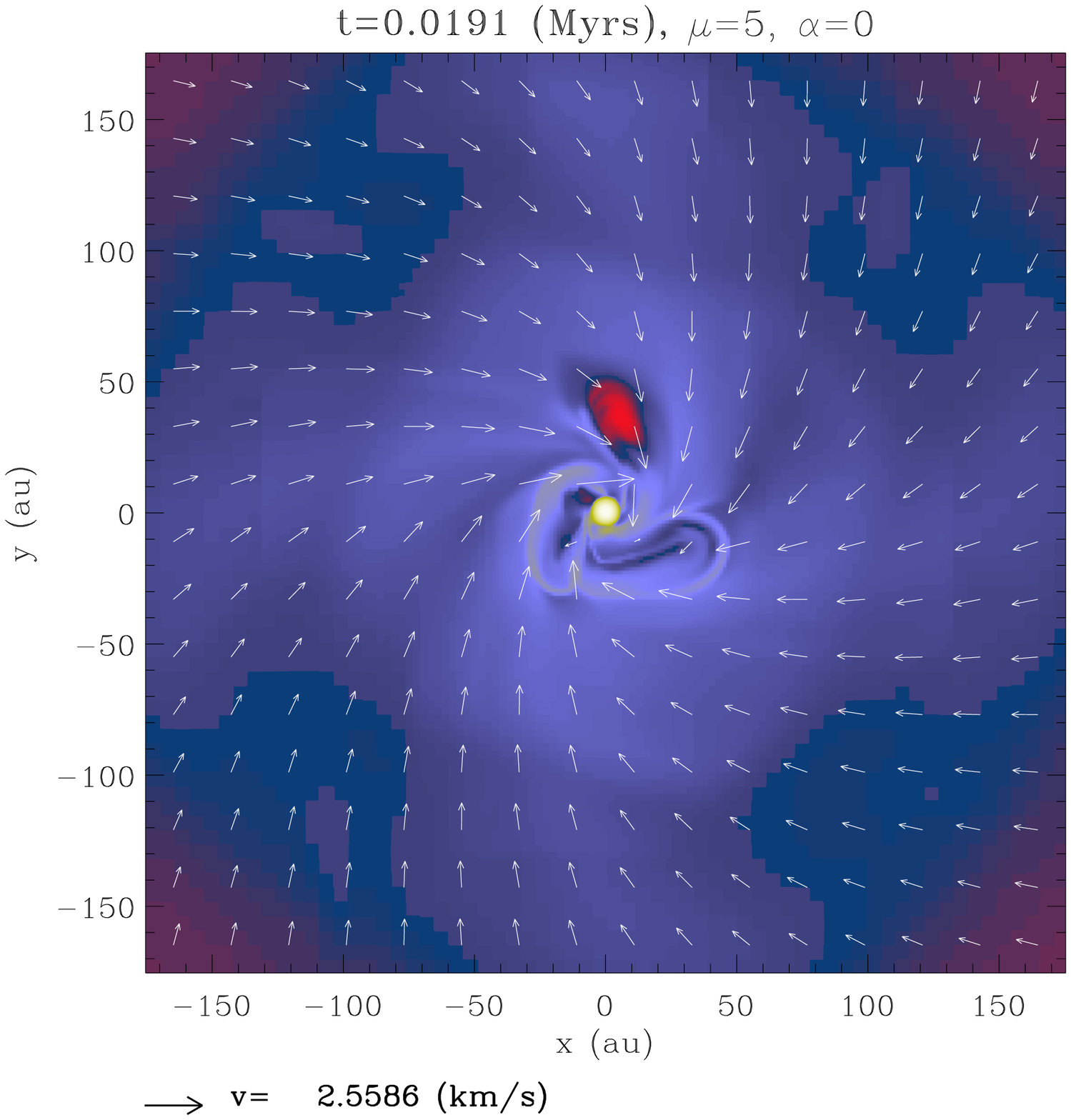}}
\put(6,0){\includegraphics[width=5.5cm]{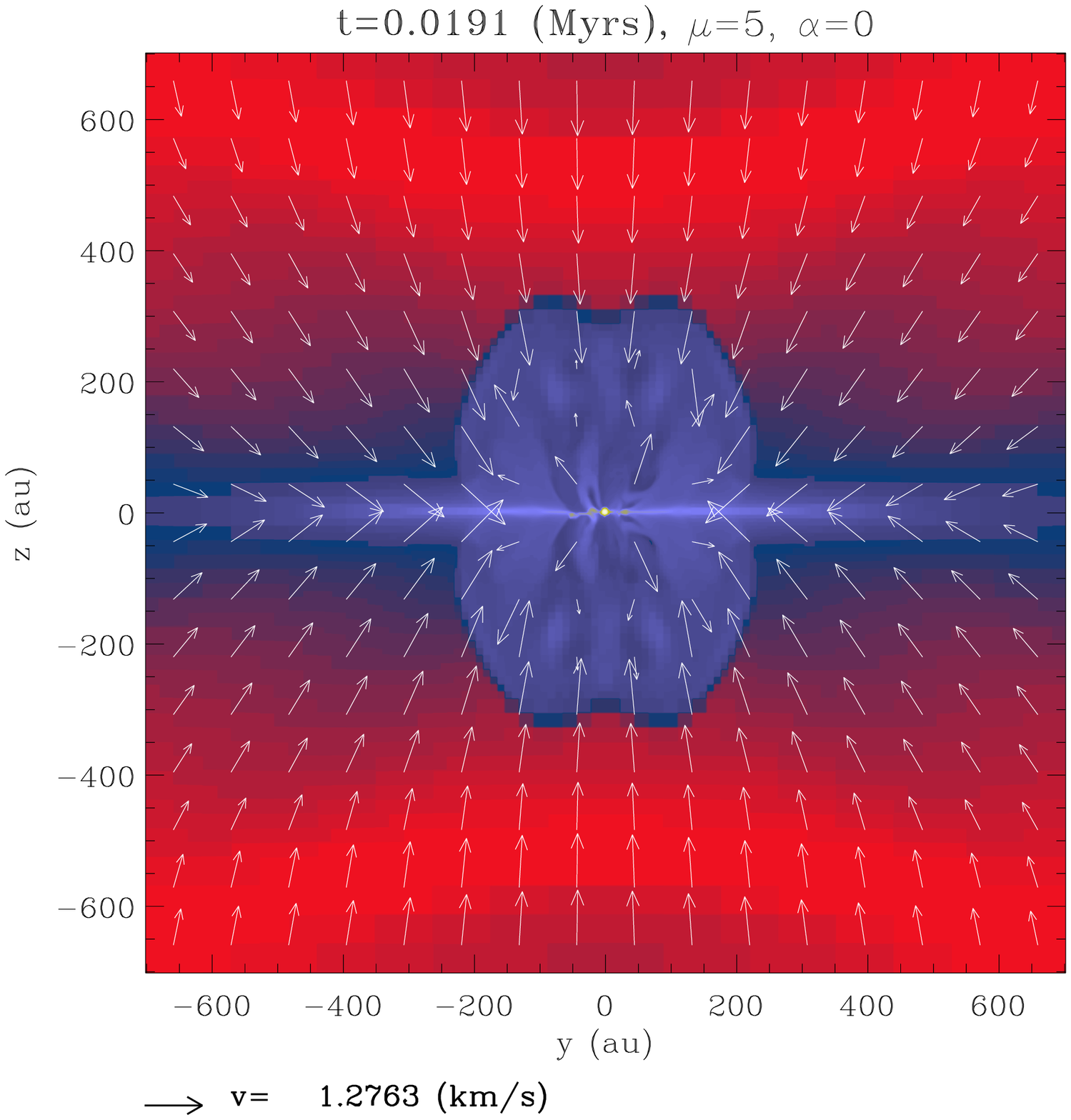}}
\put(12,5.5){\includegraphics[width=6.2cm]{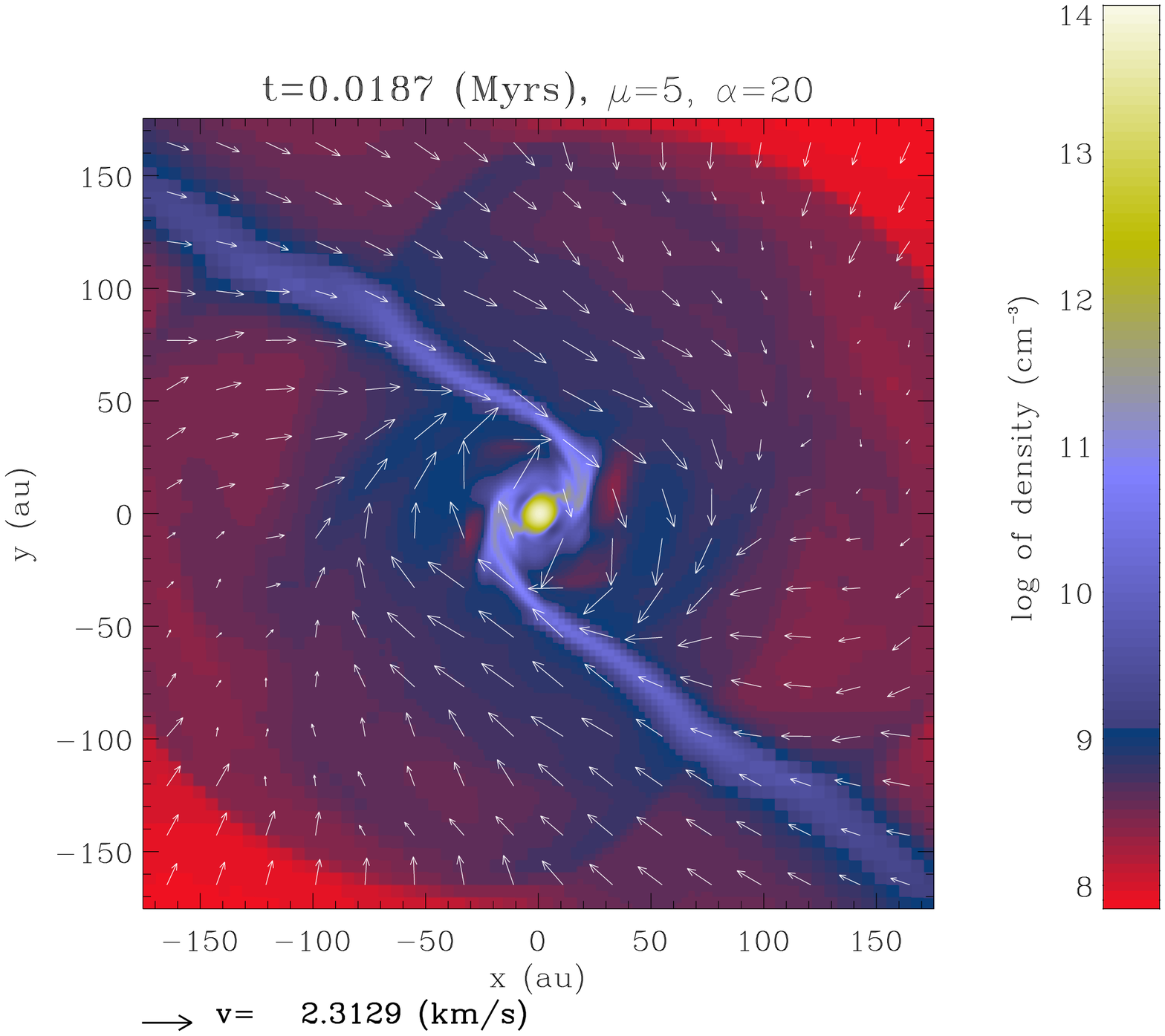}}
\put(12,0){\includegraphics[width=6.2cm]{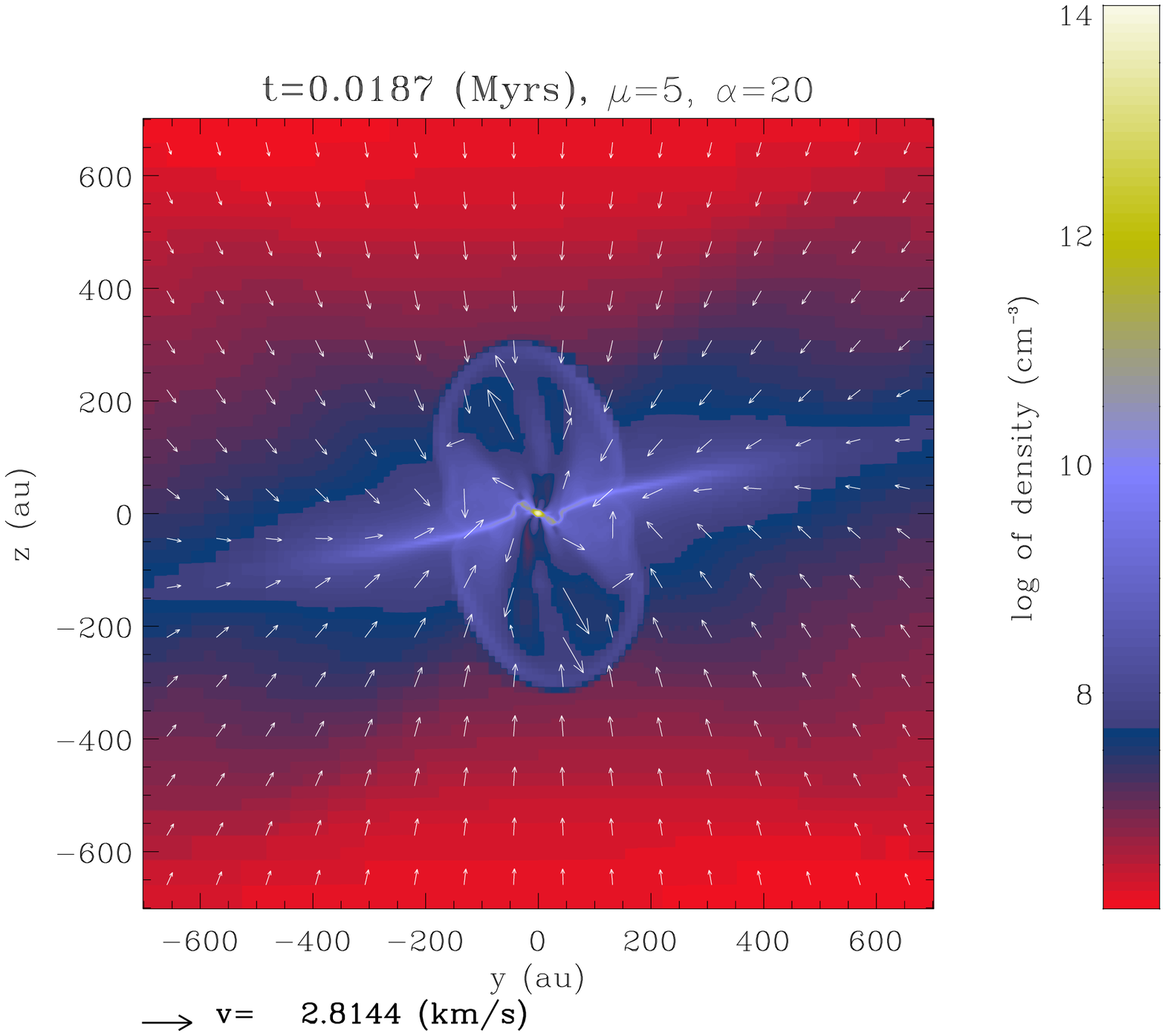}}
\end{picture}
\caption{Density and velocity profiles.  First column is for $\mu = 16$ and $\alpha=0 ^\circ$, second column is  for
$\mu=5$ and $\alpha=0 ^\circ$, while third column is for $\mu=5$ and $\alpha = 20 ^\circ$.  The first row shows the
x-y plane (therefore perpendicular to the magnetic field direction) and a zoomed image of
 the centrifugally supported disk
(when it exists). The  second row shows both the y-z plane and  the more extended magnetized pseudo-disk seen edge on ($\mid y \mid > 200$ AU). }
\label{field}
\end{figure*}

\section{Magnetic braking: physical analysis}

The classical analysis of magnetic braking, presented in Appendix, assumes that there 
is a clear distinction 
between the cloud and the intercloud medium. However, inside a collapsing core the situation differs
 a priori. 
Density gradients (typically $\rho \propto r^{-2}$) throughout the cloud make it difficult 
to discriminate between its dense inner parts and the diffuse external medium.
 It is therefore not entirely clear to what extent the classical analysis may be applied to understand
the magnetic braking operating inside a collapsing dense core. Indeed various studies have already 
demonstrated that internal magnetic braking is more complex
(e.g.; Galli et al. 2006, HF08).

A simple  analysis of the terms responsible for 
angular momentum transport through the cloud is elucidating. 
We consider a collapsing cloud permeated by a uniform magnetic field, $B_z$. 
Rotation with respect to the z-axis generates a toroidal field, $B_\theta$ 
($r$, $\theta$ and $z$ are cylindrical coordinates) and we have
\begin{eqnarray}
{v_\theta  \over \tau _{\rm brak}} \simeq  {B_z B_\theta \over 4 \pi \rho h }, \; { B_\theta  \over \tau _{\rm brak}}  \simeq {B_z v_\theta \over  h }, 
\label{brak}
\end{eqnarray}
where $\tau _{\rm brak}$ is the braking time and $h$ is the typical scale-height.  
 Combining these two equations, we obtain
\begin{eqnarray}
 \tau _{\rm brak} \simeq  {\sqrt{4 \pi \rho h^2} \over B_z }, 
\label{brak_esti}
\end{eqnarray}
which clearly shows that the braking time is shorter as, $h$, decreases.
Assuming that the column density remains roughly constant, we find that 
$\tau _{\rm brak} \propto h^{1/2}$.

As discussed 
in Galli \& Shu (1993), Li \& Shu (1996) and HF08, a magnetized cloud flattens along the field lines producing a thin magnetic pseudo-disk, which unlike centrifugally supported disks, is dynamically collapsing. We can relate the scale-height of the pseudo-disk, $h$, to the angle $\alpha$.
For an aligned system, assuming mechanical equilibrium along the z-axis, one can infer the relation (see HF08 for details)
\begin{eqnarray}
\rho_c c_s^2 \simeq B_r^2/8\pi + (\partial_z \phi)^2,
\label{equilibrium_1}
\end{eqnarray}
which reflects  that the gravity and 
magnetic pressure  compress  the gas along the z-axis while the thermal pressure resists the compression.
In this equation $B_r$, and the gravitational force $\partial_z \phi$, are estimated 
just above the disk ($z \simeq h$) while $\rho_c$ is the equatorial ($z \simeq 0$) gas density. 
As shown in HF08, for values of $\mu$ lower than 5,
the magnetic pressure produces a very significant density enhancement (typically a factor 10 or more)
 in the equatorial plane and 
very stiff gradients along the $z$ direction.
We now consider a cloud that rotates along an axis $oz'$, belonging to the x-z plane, such that
the angle, $\alpha$, between $oz$ and $oz'$  is small. We define $\Omega$ the rotation speed.
The rotation  can be decomposed into two components; one is a rotation about the  
z-axis at a velocity  $\Omega \cos(\alpha)$  and the other about the x-axis
at a (time dependent) velocity $\sim \Omega \sin(\alpha)$. As a consequence, 
the magnetic field lines are stretched. As in the aligned case, a toroidal component is created by the rotation with respect to $oz$ 
but   another component, $B_y \simeq B_z  \Omega \sin(\alpha)  \tau_{ff} $, is also generated by
 the rotation with respect to $ox$, 
where $\tau_{ff}$ is the cloud freefall time. 
Therefore, unlike  the aligned case, the 
components of the magnetic field parallel to the x-y plane  do  not vanish  in the equatorial plane.
Assuming that the angle $\alpha$  and the global cloud rotation are small,
the magnetic field remains roughly unchanged with the exception of  its y-component. 
The mechanical equilibrium equation becomes
\begin{eqnarray}
\rho_c c_s^2 + B_y^2/8\pi \simeq B_r^2/8\pi + (\partial_z \phi)^2. 
\label{equilibrium_2}
\end{eqnarray}
where $B_r$ is the radial field  corresponding to the case without rotation.
Equation~\ref{equilibrium_2} shows that the $B_y$ component generated by the cloud rotation limits the magnetic compression along the z-axis.
Therefore, the typical width of the pseudo disk, $h$, increases as $\alpha$ increases, and the efficiency 
of the magnetic braking ($\propto h^{-1/2}$) is consequently reduced.
In addition, since the rotation axis is not parallel to the z-axis, the rotation itself will also tend
 to impede the contraction along the z-direction.

\setlength{\unitlength}{1cm}
\begin{figure}
\includegraphics[width=7cm]{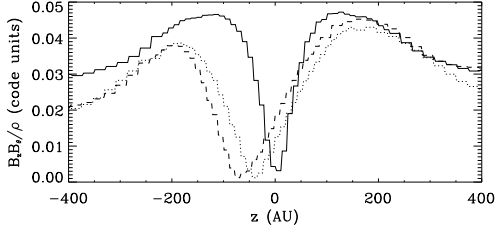}
\caption{$|B_z B_\theta / \rho|$ (see Eq.~\ref{brak}) as a function of $z$ at $y=300$ AU. Solid line is $\theta=0^\circ$, 
dotted line $\theta=10^\circ$ and dashed line $\theta=20^\circ$.}
\label{brak_estim}
\end{figure}

\setlength{\unitlength}{1cm}
\begin{figure*}[t]
\begin{picture}(0,5)
\put(0,0){\includegraphics[width=5.5cm]{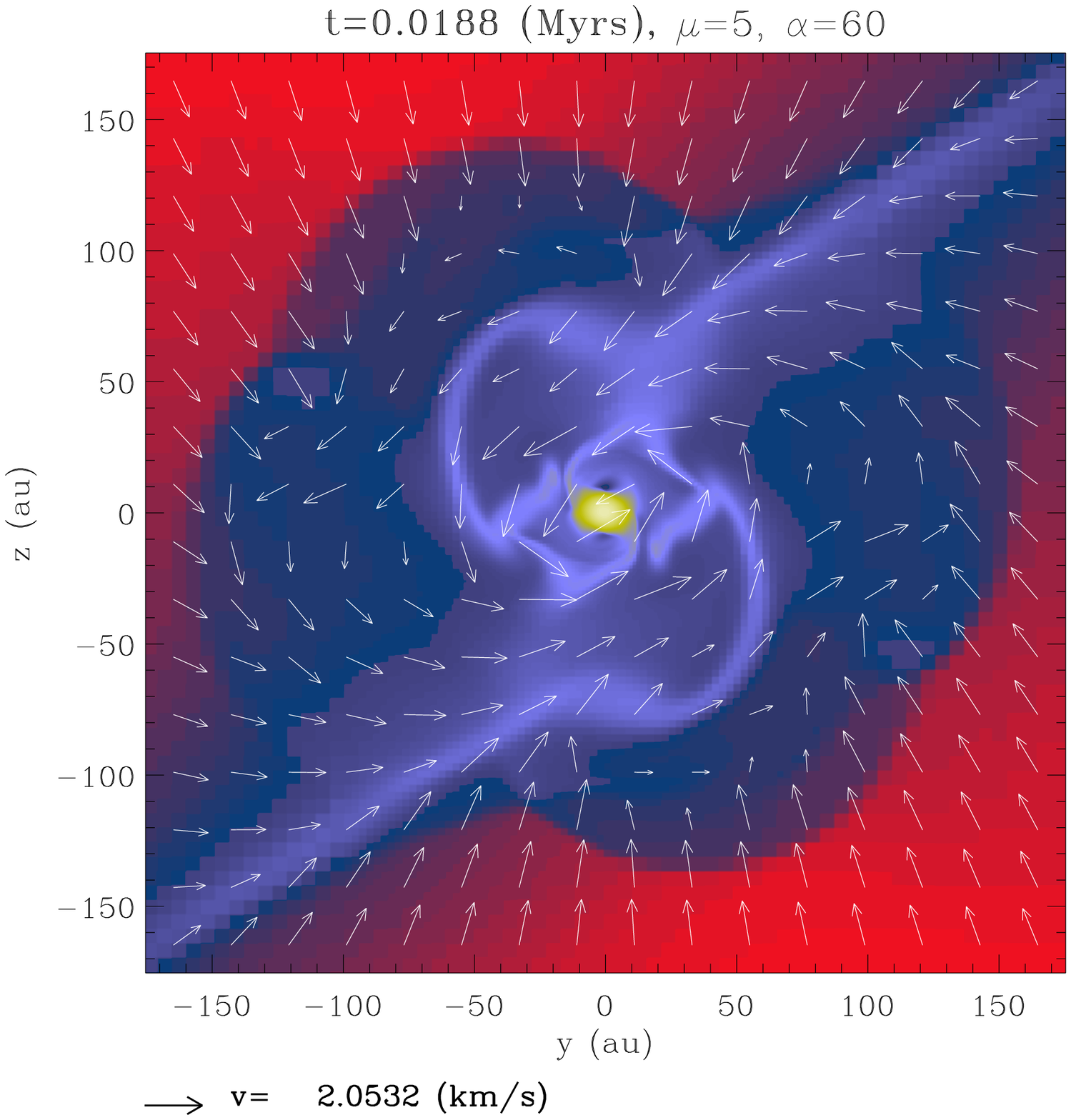}}
\put(6,0){\includegraphics[width=5.5cm]{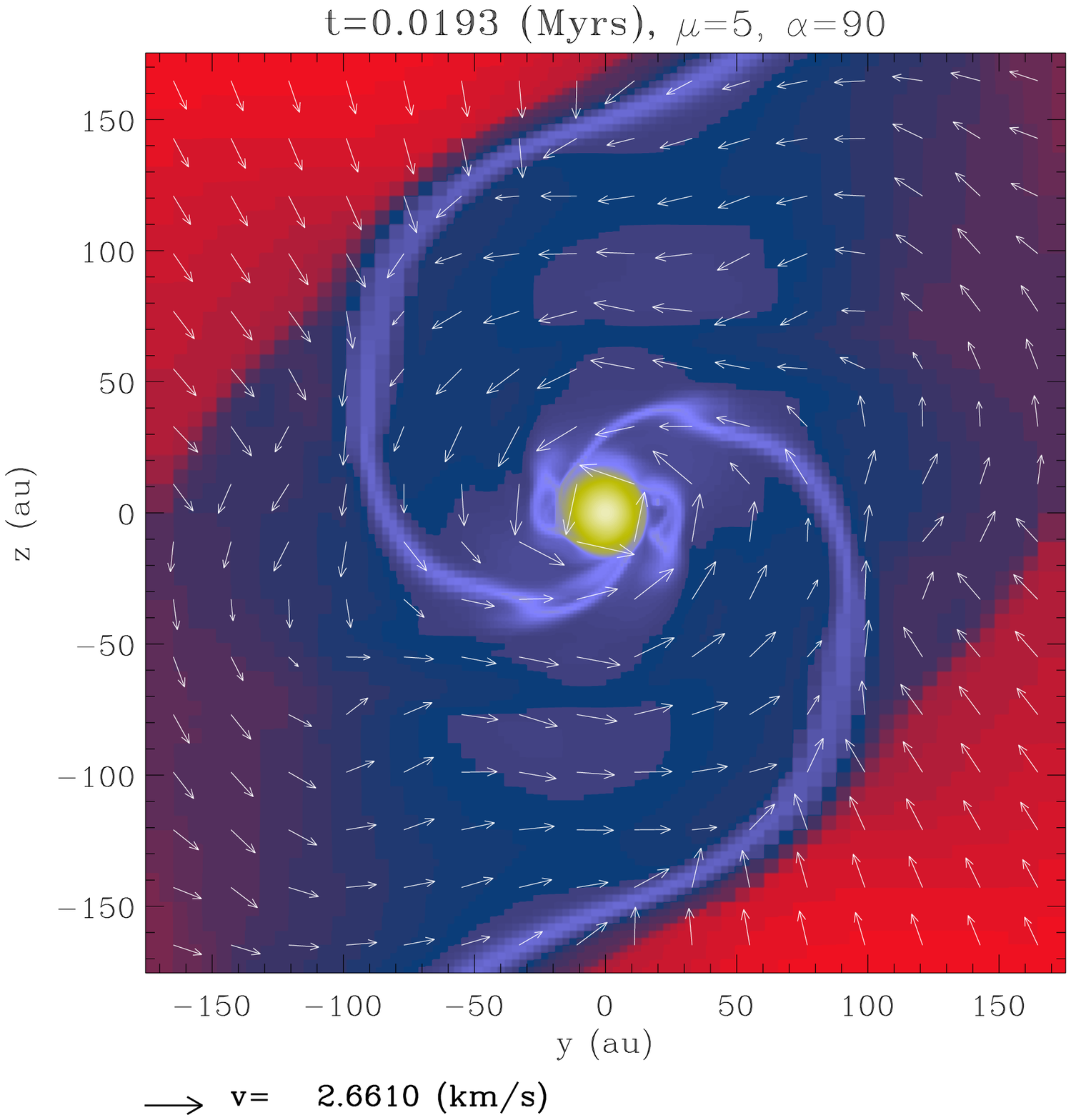}}
\put(12,0){\includegraphics[width=6.2cm]{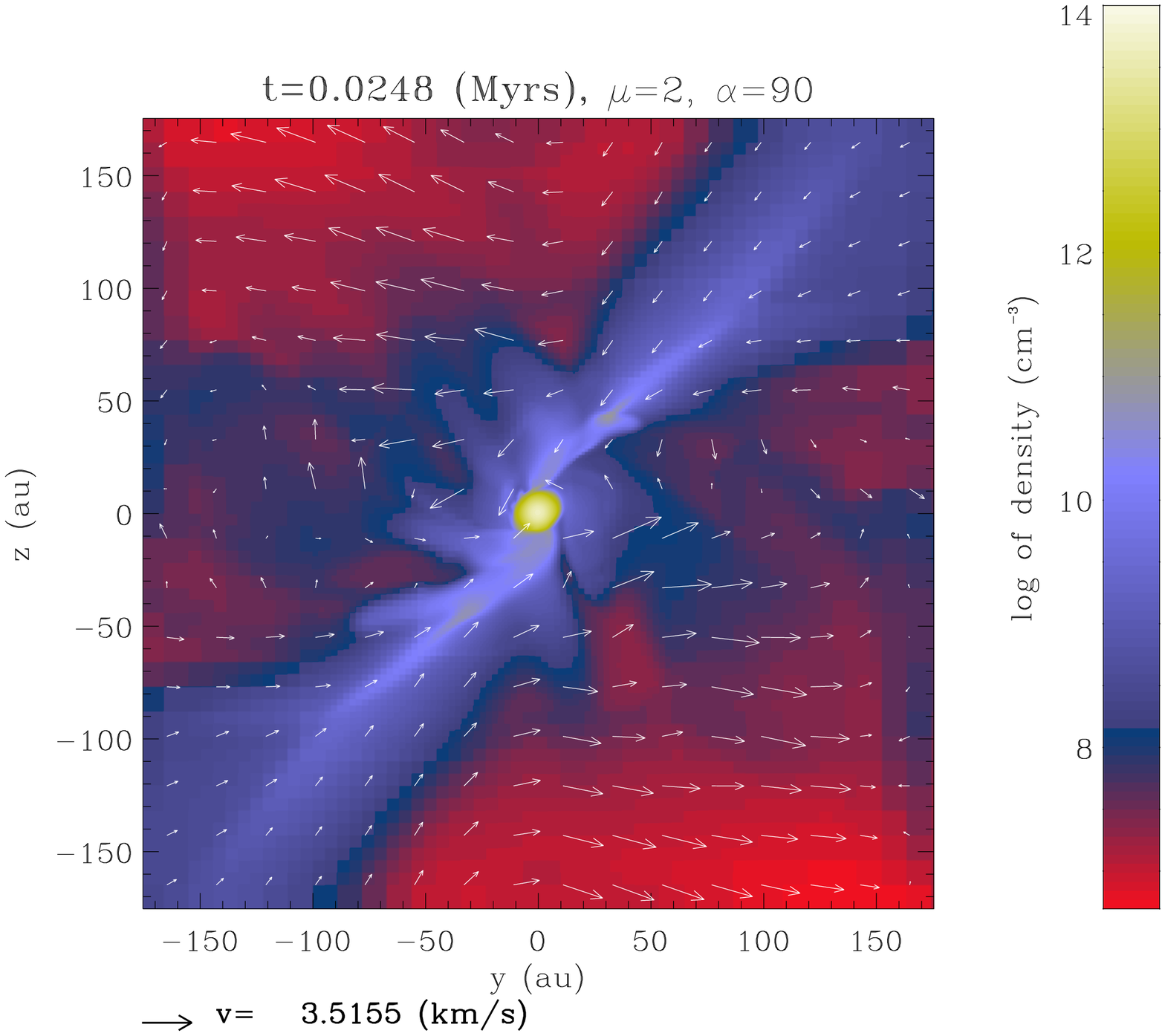}}
\end{picture}
\caption{Density and velocity cut through the y-z plane. The centrifugally supported disk, when it exists, 
is seen nearly face on.
First panel is for $\mu=5$ and $\alpha=60 ^\circ$, 
second panel is for $\mu=5$ and $\alpha=90 ^\circ$ while third panel is for 
$\mu=2$ and $\alpha=90 ^\circ$.}
\label{field2}
\end{figure*}

\section{Numerical simulations}

\subsection{Numerical setup}
To carry out our numerical simulations, we used the AMR code Ramses (Teyssier 2002, Fromang et al. 2006).
Throughout the simulations, 
the Jeans length is resolved with at least 10  cells. An HLLD solver is employed.
The initial conditions consist of a  1 solar mass spherical cloud whose profile, given by $\rho(r) =  \rho_c / (1 + (r/r_0)^2)$, resembles the observed cores.  
We impose a density contrast of 10 between the central density $\rho_c=8 \times 10^6$ cm$^{-3}$
and the edge density, $\rho_e$. Outside the cloud, the gas has a 
density of $\rho_e / 10$ and it is in pressure equilibrium with the cloud edge.
As in HF08, we use a barotropic equation of state.
The cloud is initially in solid body rotation and threaded by a  
magnetic field, along the z-axis, whose intensity is proportional to the total column density through the cloud.
The rotation axis is in the x-z plane and at an angle $\alpha$ with respect to the magnetic field.
The ratio of rotation to gravitational energy is equal to $\beta \simeq 0.03$, while the degree
of magnetization is determined by the parameter $\mu$. The initial thermal to
 gravitational energy  ratio is $\sim 0.25$.

\subsection{Results}
For various values of $\mu$ and $\alpha$, Fig.~\ref{field} shows
the density and velocity distribution in the x-y plane 
(first row) and y-z plane (second row). These correspond to the initial direction 
of the magnetic field being respectively
 perpendicular and parallel to the plane of the snapshot. Note that the rotation axis is also 
either perpendicular or parallel, except in the case of  
 $\alpha=20 ^\circ$  for which it is slightly tilted.
At the time of the snapshots, the mass of the gas denser than $10^{10}$ cm$^{-3}$ (the density at which the gas becomes 
adiabatic) is $\sim 0.15$ M$_\odot$, while the mass of the gas denser than $10^{13}$ cm$^{-3}$ is $\sim 0.1$ M$_\odot$.

The first two columns show results for  the cases $\mu=16$ and $\mu=5$, respectively, with $\alpha=0^\circ$.
In the $\mu=16$ case, a dense disk whose diameter is $\sim 200$ AU forms 
around the central object. The size of the disk is estimated by determining the position at which the velocity is 
roughly perpendicular to the disk radius, so that the rotation velocity dominates over the infall motions.
The cut across the y-z plane also shows an outflow being launched perpendicularly to the disk. 
In the $\mu=5$ case, no disk of significant size forms.
The gas instead falls directly towards the  central thermally supported core. 
The entire configuration is unstable leading to  significant departures from the axisymmetric configuration.
The cut
across the y-z plane also shows the thin pseudo-disk ($\mid y \mid > 200$ AU, $\mid z \mid \simeq 100$ AU for $\rho=10^8$ cm$^{-3}$), 
which is compressed by the magnetic pressure exerted by the curved field lines.
The top panel of the third column ($\mu=5$, $\alpha=20^\circ$) shows that a 
disk-like structure of $\sim150$ AU forms around the
central object (as shown by the velocity field, rotation dominates over infall in this structure).
We note that  the structure appears to be non-axisymmetric  and that the entire velocity field is 
more complex than in the first case ($\mu=16$, $\alpha=0^\circ$). 
For $\mu=5$ with $\alpha=20^\circ$ (bottom panel of third column) the pseudo disk, visible at $\mid y \mid > 200$ AU, is twisted and overall thicker by a factor of $\sim 2-3$ than the case $\mu=5$, $\alpha=0^\circ$ (bottom panel of the second column). Additionally, we have performed simulations of the case $\alpha=10^\circ$ for which we also find a disk. According to the analysis presented 
in sect.~2, we believe that the thickness of the pseudo-disk is at the origin of the formation of 
a rotationally dominated disk-like structure, since it reduces the magnetic braking efficiency as 
shown by Eq.~(\ref{brak}).
To confirm that the different behaviours are indeed related to  differences in the 
 magnetic braking efficiency, we plot in Fig.~\ref{brak_estim} the quantity $|B_z B_\theta| / \rho$ which appears in Eq.~(\ref{brak_esti}). The cases $\theta=0^\circ$ (solid line),
 $\theta=10^\circ$ (dotted line) and  $\theta=20^\circ$ (dashed line) are shown. As can be seen, the quantity $|B_z B_\theta| / \rho$ is higher, and the two peaks are closer (equivalently the gradients between the peaks are stiffer), for the aligned case; implying that 
the quantity $B_z B_\theta / (\rho h)$ is higher by at least $\sim 50-100 \%$ than for the two misaligned configurations ($B_\theta =0$ at $z=0$, for symmetry reasons). 
This sudden change of behaviour as $\alpha$ increases, indicates,
 as suggested by  Eqs.~(\ref{equilibrium_1}-\ref{equilibrium_2}), that the 
aligned configuration is probably {\it too restrictive} for studying disk formation in collapsing magnetized cores.

Figure~\ref{field2} shows the density and velocity distributions in the y-z plane for larger values of 
$\alpha$. The first panel illustrates the $\mu=5$ and $\alpha=60 ^\circ$ case, while the second panel 
shows results for $\alpha=90 ^\circ$. In both cases a disk of $\sim150-200$ AU forms.
We note that the densities in these disks are significantly lower than those observed in the
 $\mu=16$ and 
$\alpha=0^\circ$ case (first row of Fig.~\ref{field}); their structure is  clearly
more  complex, and we note the presence of extended spiral arms, which are almost one order
of magnitude denser than in the rest of the disk.
Finally, the third panel of Fig.~\ref{field2} shows results for 
a highly magnetized cloud rotating perpendicularly to the magnetic field ($\mu=2$, $\alpha=90^\circ$). 
In this case, even though the magnetic braking is less efficient than in the 
aligned configuration,  the magnetic field is so strong that no disk forms.
Interestingly, we also see that a relatively strong outflow is launched.
For completeness, we  also considered the case ($\mu=3$, $\alpha=90 ^\circ$)
 for which we find that a small disk-like object 
of size 50-100 AU forms.
Thus, we conclude that in the perpendicular case and within the framework 
of ideal MHD, disks can form for values of $\mu$  higher than $\simeq$3. It seems probable
 to us that, given 
the complex field geometry induced in these non aligned configurations, ambipolar diffusion as well as ohmic dissipation 
may play  a role in the formation of small disks.

\section{Conclusion}
To study the formation of circumstellar disks, we have performed numerical simulations 
of the gravitational collapse of magnetized 
dense prestellar cores for a range of values of $\mu$
and $\alpha$ (the initial angle between magnetic field and the rotation axis). Our main conclusions are the following. In the aligned case ($\alpha=0^\circ$) disk formation 
is prevented for values of $\mu$ as high as $\sim 5-10$. In the non aligned cases, and even 
for relatively small values of $\alpha \simeq 10-20^\circ$, disks can form 
for smaller values of $\mu \simeq 4-5$, corresponding to larger magnetic field intensities.
The main reason for this change of behaviour is a decrease, for increasing $\alpha$, of the magnetic braking efficiency, which is linked to an increase in the thickness of the magnetized pseudo-disk. 
This is caused by the magnetic field lines being twisted by the 
rotation about  an axis {\it parallel} to  the plane of the pseudo-disk.
When $\alpha=90 ^\circ$, we find that disks may form for smaller values of $\mu$, as long as $\mu > 2-3$, and for even lower values of $\mu$, disk formation does not seem to be possible.
Nevertheless, for these highly magnetized configurations, 
the question of  whether a disk may form at later times,
or because of non-ideal MHD effects (Hosking \& Whitworth 2004, Machida et al. 2008, Mellon \& Li 2009), 
remains unanswered.
We recall that although Belloche et al. (2002) observe a significant amount of rotation in the envelope of 
IRAM04191, 
they exclude a disk of size larger than 20 AU.  This result is very difficult to explain 
without a magnetic field, but agrees at least qualitatively with  our conclusion if the source 
is sufficiently magnetized (Troland \& Crutcher 2008).

Another important point is that while in the purely hydrodynamical case, and for the cloud parameters used, 
fragmentation  occurs; we do not identify fragmentation in any of the magnetized cases explored.
It is however possible that with a better treatment of the thermal processes, the magnetized disks 
which form in the low magnetized configurations, may also fragment.

\section{Acknowledgments}
This work was performed using HPC resources from GENCI-cines (2009-042036) and cemag. AC is supported
 by the ANR grant MAGNET and by the Marie Curie Reintegration Grant MAGPLUS.
We thank Shantanu Basu 
as well as an anonymous referee for comments which  have improved the paper significantly.

\appendix

\section{Classical analysis of the magnetic braking: cloud embedded into an external medium}

In the context of interstellar clouds, the classical analysis of magnetic braking 
(e.g.; Mouschovias 1991, Shu et al. 1987), considers a rigid and dense axisymmetric cloud. 
We define $\rho_c$ 
to be its density, $R$ its radius and  $Z$ its height.
The cloud is surrounded by a diffuse inter cloud medium of density $\rho_{ic}$. 
The typical timescale for magnetic braking $\tau_{br}$, corresponds to the time necessary for torsional Alfv\'en waves
induced by the twisting of the magnetic field line, to propagate over a distance $l$, such that 
the mass of gas swept by the waves is comparable to the mass of the cloud itself. At this point, 
 a significant fraction of the cloud angular momentum has been transferred to the intercloud medium.

Two cases can be considered. First, when the magnetic field and the rotation axis are aligned, the waves propagate along the magnetic field at the Alfv\'en speed $V_a$, 
 and $l \rho_{ic} \simeq Z \rho_c$, leading to
\begin{eqnarray}
\tau_{br} \simeq { Z  \over V_a} { \rho_c \over \rho_{ic} }. 
\label{brake_align}
\end{eqnarray}
The second case corresponds to the magnetic field and the rotation axis being perpendicular 
to each other, and the waves propagating in the equatorial plane of the cloud.
The intercloud medium, which at time $t=\tau_{br}$ is reached by the torsional Alfv\'en waves,
is located in a cylinder of radius $l$ and height $Z$. In this case,
$( (\tau_{br}  V_a)^2-R^2) \rho_{ic} \simeq R^2  \rho_c$, which gives
\begin{eqnarray}
\tau_{br}   \simeq {R \over V_a}  \sqrt{ { \rho_c  \over \rho_{ic}} + 1}.
\label{brake_perp}
\end{eqnarray}

Since in typical astrophysical circumstances the intercloud medium has a density 
that is low with respect to the cloud density, Eqs.~(\ref{brake_align}) and 
(\ref{brake_perp}) show that the braking is  usually more efficient 
when the magnetic field is perpendicular to the rotation axis than when 
it is parallel. However, this conclusion is obviously correct only 
when $R \simeq Z$, i.e., if the cloud aspect ratio  is not too different from 1.
In particular, from Eqs.~(\ref{brake_align})-(\ref{brake_perp}) we see
that if $Z/R \ll 1/\sqrt{\rho_c/\rho_{ic}}$, the braking time is shorter
in the aligned case than for the perpendicular configuration.

Previous studies have demonstrated that magnetized clouds are
usually very flat because of the magnetic compression exerted by the radial 
component of the magnetic field (see e.g.; Li \& Shu 1996, HF08).
These magnetized sheets, which are called pseudo-disks, are perpendicular to 
the average magnetic field. 
In the same way, centrifugally supported disks are also very flat objects
that are perpendicular to the rotation axis. For these 
extreme configurations, the magnetic braking time can obviously be longer
when the magnetic field and the rotation axis are perpendicular than when they are parallel (depending 
on the respective values of $R$, $Z$ and $\rho_c/\rho_{ic}$).
Note that strictly speaking, if the magnetic field and the rotation axis are not aligned
with each other, the 
resulting structure is fully tridimensional rather than axisymmetric.

 It is interesting to compare Eqs.~(\ref{brake_align})-(\ref{brake_perp}) with
Eq.~(\ref{brak_esti}). Even though the latter is identical to the first  for 
$\rho_c=\rho_{ic}$,  there are two major differences. First, even when  $R \simeq Z$, 
the magnetic braking is not significantly more efficient in the perpendicular configuration 
than in the aligned one.  The relative efficiency of the magnetic braking in the 
two configurations is, instead, directly proportional to the cloud aspect ratio. 
Second, as the cloud is compressed 
along the field lines, the quantity $Z / \sqrt{\rho_c}$   is simply proportional to $Z^{1/2}$, implying 
that the magnetic braking time in the aligned configuration decreases. Again this is unlike the case of a rigid cloud
embedded in a diffuse intercloud medium.

\end{document}